\documentclass[twocolumn,showpacs,preprintnumbers, superscriptaddress, prl]{revtex4-1}
\usepackage{amsmath}
\usepackage{stmaryrd}
\usepackage{txfonts}
\usepackage{amssymb}
\usepackage{mathrsfs}
\usepackage{graphicx}
\usepackage{dcolumn}
\usepackage{bm}
\usepackage{epsfig}
\usepackage{color}
\usepackage{hyperref}

\begin{document}
\preprint{\href{http://dx.doi.org/10.1103/PhysRevB.87.214511}{S.-Z. Lin and A. E. Koshelev, Phys. Rev. B {\bf 87}, 214511 (2013).}}

\title{Linewidth of the electromagnetic radiation from Josephson junctions
near cavity resonances}
\author{Shi-Zeng Lin}
\email{szl@lanl.gov}
\affiliation{Theoretical Division, Los Alamos National Laboratory, Los Alamos, New Mexico
87545, USA}
\author{Alexei E. Koshelev}
\email{koshelev@anl.gov}
\affiliation{Materials Science Division, Argonne National Laboratory, Argonne, Illinois
60439, USA}

\begin{abstract}
The powerful terahertz emission from intrinsic Josephson junctions in
high-$T_c$ cuprate superconductors has been detected recently. The
synchronization of different junctions is enhanced by excitation of the
geometrical cavity resonance. A key characteristics of the radiation is its
linewidth. In this work, we study the intrinsic linewidth of the radiation
near the internal cavity resonance. Surprisingly, this problem was never
considered before neither for a single Josephson junction nor for a stack of
the intrinsic Josephson junctions realized in cuprate superconductors. The
linewidth appears due to the slow phase diffusion, which is
determined by the dissipation and amplitude of the noise. We found that both
these parameters are resonantly enhanced when the cavity mode is excited but
enhancement of the dissipation dominates leading to the net suppression of
diffusion and dramatic narrowing of the linewidth. The line shape
changes from Lorentzian to Gaussian when either the Josephson frequency shifted away from the resonance or the temperature is increased.
\end{abstract}

\date{\today}
\maketitle

In a Josephson junction (JJ) biased by a dc voltage $V$ the supercurrent
oscillates with the angular frequency $\omega_J =2eV/\hbar $. This allows to
use the JJs as high-frequency electromagnetic (EM) generators. The radiation from a single JJ however is weak, only several picowatts. The radiation power can be enhanced using arrays of JJs. \cite{Jain84,Barbara99} In 2007, coherent and strong terahertz (THz) radiations from intrinsic Josephson junctions
(IJJs) of $\mathrm{{Bi_{2}Sr_{2}CaCu_{2}O_{8}}}$ (BSCCO) \cite{Kleiner92} has been observed
experimentally \cite{Ozyuzer07}. In this experiment, the radiation power was
estimated as $0.5\ \mathrm{\mu W}$ which is several orders of magnitude stronger
than that from a single junction. The frequency has ranged from $0.4$ to $0.8$ THz and inversely  proportional to the mesa width. Such an observation has led the authors of
Ref. \onlinecite{Ozyuzer07} to conclusion that the strong radiation is due to the
excitation of cavity modes insides the mesa.

 Significant progress has been made in
the last several years \cite{Bulaevskii07,kadowaki08,Wang09,Wang10,Tsujimoto10,Kashiwagi2012,Tsujimoto12b,Benseman11,Orita2010,szlin08,szlin08b,Koshelev08b,szlin12a,An2013} and the radiation power is enhanced by two orders of magnitude. Recently, much attention has
been paid to achieve frequency tunability \cite{Benseman11,Tsujimoto12b} and to
enhance radiation power by using mesa arrays. \cite{Orita2010} These developments suggest that the IJJs in high-$T_{c}$ superconductors are
extremely promising for development of efficient sources of THz EM waves.
Such sources would have wide applications in different areas such as medical
imaging, security, and new spectroscopy where progress is limited by lack of
compact solid state generators. \cite{Tonouchi07}

Besides the radiation power, another figure of merit of the THz radiation is the linewidth. The linewidth from a single point junction has been investigate about half century
ago \cite{Larkin1968,Stephen68,Dahm69}. Fluctuations of Cooper pairs
\cite{Stephen68} and later the fluctuations of quasiparticles \cite{Dahm69} are
taken into account in theoretical calculations of the linewidth and a satisfactory
consistency between theory and experiments was achieved. The radiation linewidth
from a tall stack of IJJs was calculated in Ref. \onlinecite{Bulaevskii11} assuming
that the main source of damping is coming from external radiation. An extremely
narrow relative linewidth (defined as the ratio of linewidth to radiation frequency)
of order $10^{-9}$ was obtained. The line shape of radiation
coming from BSCCO mesas at cavity resonance has been measured recently.
\cite{Kashiwagi2012,LiLinewidth12} The narrowest lines with width $~20$MHz are found
in the high bias regime \cite{LiLinewidth12} while at low bias regime a typical
linewidth is about $0.5$ GHz. \cite{Kashiwagi2012,LiLinewidth12} As excitation of the cavity mode is essential for synchronization of IJJs, it is important to understand its influence of the radiation line shape. Surprisingly, no theory exists on the linewidth of the radiation from JJs near cavity resonances neither for a long JJ nor for a stack of IJJs.

Here we present both analytical and numerical study on the linewidth of high
frequency radiation from a JJ or a stack of IJJs near cavity resonances due
to thermal fluctuations. The linewidth broadening is caused by the diffusion
of the phase at wavenumber $\mathbf{k}=0$. The line shape changes from
Lorentzian to Gaussian when temperature is increased. Fluctuations with nonzero wave vectors lead to the suppression of the radiation power. As voltage is tuned
close to the cavity resonance, the line width is sharpening significantly
and being inverse proportional to the volume of the system. For typical
parameters, the line shape is Lorentzian and the linewidth can be expressed
in terms of $IV$ characteristics. We give an theoretical limit for the linewidth using typical parameters for BSCCO.

For simplicity, let us first consider a single JJ with spatial modulation of the
critical current \cite{Koshelev08}. The modulation of the critical current may be
due to the defects in the junction, also can be introduced intentionally. A single
junction with uniform external magnetic fields and the $\pi$ phase kink state in a
stack of IJJs also reduce to this model \cite{Koshelev10}. The equation of motion in
dimensionless units can be written as \cite{BaroneBook,Koshelev08}
\begin{equation}  \label{eq1}
\partial _t^2\theta + \beta {\partial _t}\theta + g(x)\sin\theta -
\nabla_{2d}^2\theta = J_n(\mathbf{r},t) + {J_{\mathrm{ext}}},
\end{equation}
where $\beta$ is the damping do to the quasiparticle conductivity, $
\nabla_{2d}^2\equiv\partial_x^2+\partial_y^2$ and $\mathbf{r}=(x, y)$. The
radiation is weak and the boundary condition can be approximated as $
\partial_\mathbf{n}\theta=0$, where $\mathbf{n}$ is a unit vector normal to
the surface. The spatial modulation is assumed along the $x$ direction. $J_n(
\mathbf{r},t)$ is the white-noise current satisfying the fluctuation dissipation
theorem (FDT) valid in equilibrium
\begin{equation}  \label{eq2}
\langle J_n\rangle=0, \ \ \ \langle J_n(\mathbf{r}, t)J_n(\mathbf{r}^{\prime
},t^{\prime }) \rangle=2 T \beta\delta(t-t^{\prime })\delta(\mathbf{r}-
\mathbf{r}^{\prime }).
\end{equation}
When the JJ is driven into the voltage state where the phase rotates
following the ac Josephson relation, the FDT is violated as demonstrated
below.

\begin{figure}[t]
\psfig{figure=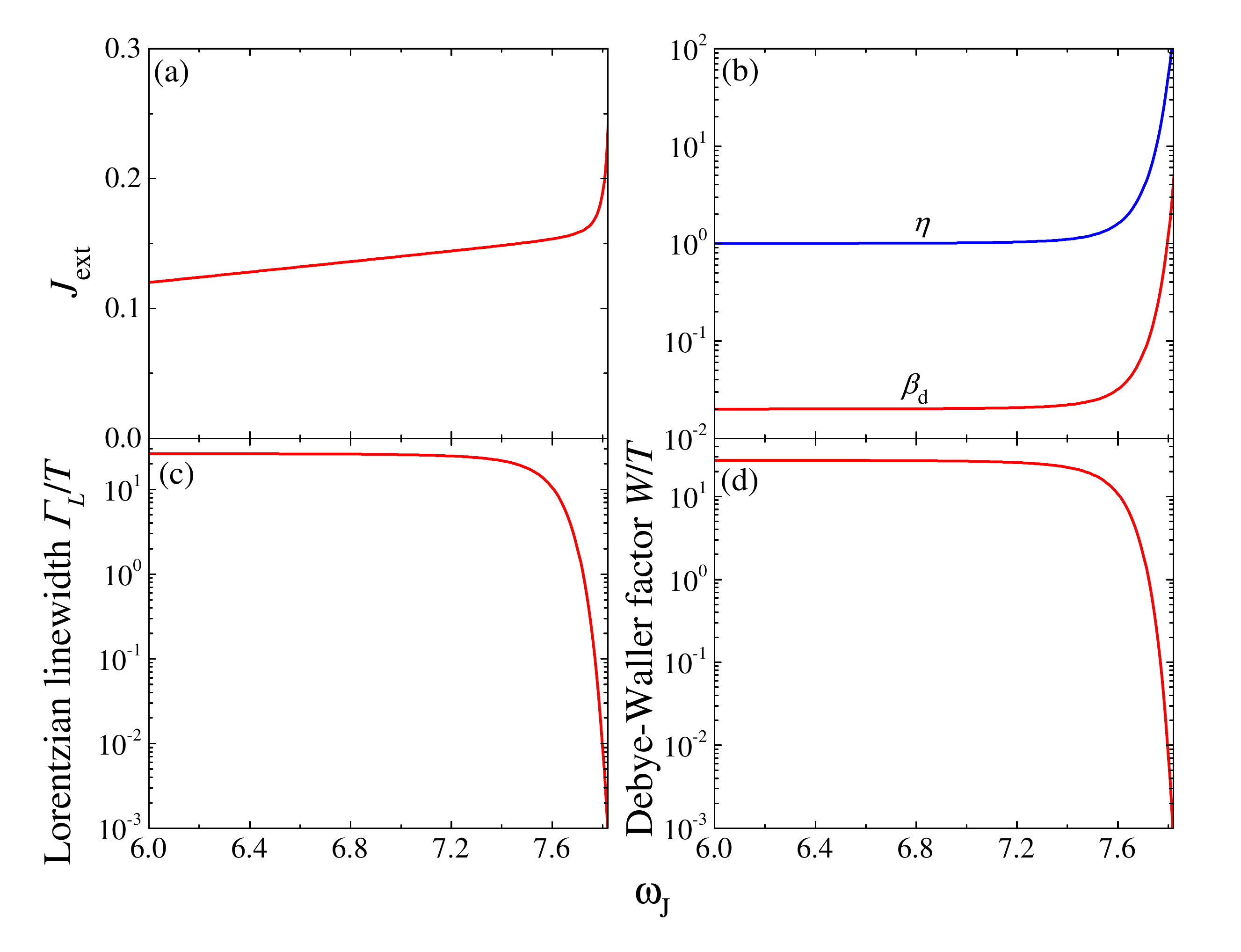,width=\columnwidth}
\caption{(color online) (a) $IV$ curve (b) the dynamic conductivity $\beta_d$ and the inertial term
$\protect\eta$ (c) the Lorentzian linewidth, and (d) the
Debye-Waller factor normalized by temperature $T$. Parameters are: $\protect\beta=0.02$,
$L_x=0.4$, $L_y=1.5$ and $F=\protect\pi/4$ for a step modulation of critical
current $g(x)=-\mathrm{sign}(x-L_x/2)$.}
\label{f1}
\end{figure}

The Josephson junction is characterized by the intrinsic cavity modes with wavenumbers $\mathbf{k}_{nm}=(k_{xn},\ k_{ym})=(n\pi /L_{x},\ m\pi /L_{y})$ and frequencies $\omega_{nm}=\sqrt{(n\pi /L_{x})^2+(m\pi /L_{y})^2}$.
The cavity mode may be selected by the voltage of the junction, which determines the
Josephson frequency.
The $x$-modulated Josephson current couples the Josephson oscillations to the cavity
modes with wavenumbers $\mathbf{k}_{n0}$.
Without loss of generality, we consider the thermal
fluctuations around the mode $(\pi /L_{x},0)$. In the voltage state without noise
$J_{n}=0$, the phase is described by $\theta _{0}=\omega_J t+\mathrm{Re}[A\exp
(i\omega_J t)]$ with $ A=iF/(-\omega_J ^{2}+i\beta \omega_J +k_{x1}^{2})$ and
$F=\frac{2}{{L_{x}}} \int_{0}^{L_{x}}dx\cos (k_{x1}x)g(x)$. Here
$\omega _{J}$ is the angular frequency determined by the dc voltage $V$, $\omega_J=V$. We restrict to the
analytically tractable region $A\ll 1$. In this case, the $IV$ curve $J_{\mathrm{ext}}=\omega_J
t+\langle \sin \theta _{0}(\mathbf{r},t)\rangle _{\mathbf{r},t}$ is given by
\begin{equation}
J_{\mathrm{ext}}=\beta \omega_J +\frac{{F^{2}}}{4}\frac{{\beta \omega_J }}{{{{({
\omega_J ^{2}}-k_{x1}^{2})}^{2}}+{\beta ^{2}}{\omega_J ^{2}}}},  \label{eq3}
\end{equation}
where $\langle \cdots \rangle _{\mathbf{r},t}$ is the spatial and temporal average. We introduce
the dynamic conductivity $\beta _{d}\equiv dJ_{\mathrm{ext} }/d\omega_J $
\begin{equation}
\beta _{d}=\beta +\frac{\beta F^{2}}{4}\frac{\left( \omega_J
^{2}-k_{x1}^{2}\right) ^{2}-\beta ^{2}\omega_J ^{2}-4\omega_J ^{2}\left( \omega_J
^{2}-k_{x1}^{2}\right) }{\left[ \left( \omega_J ^{2}-k_{x1}^{2}\right)
^{2}+\beta ^{2}\omega_J ^{2}\right] ^{2}}.  \label{eq4}
\end{equation}
The first part is due to the usual conductivity $\beta $ and the second part is due
to resonant contribution, which sharply increases as $\omega_J \rightarrow k_{x1}$. As
will reveal later, the linewidth is determined by $ \beta _{d}$. In Fig. \ref{f1},
the typical $IV$ curve and $\beta _{d}$ are shown, both of which are enhanced at the
resonance.

To calculate the linewidth, we need to know the response of the phase to the noise
current. The phase is $\theta =\theta _{0}+\tilde{\theta}$, with the phase due to
the noise $\tilde{ \theta}$ being governed by
\begin{equation}
\partial _{t}^{2}\tilde{\theta}+\beta {\partial _{t}}\tilde{\theta}+g(x)\cos
(\theta _{0})\tilde{\theta}-\nabla _{2d}^{2}\tilde{\theta}=J_{n}.
\label{eq5}
\end{equation}
Phase diffusion is determined by slow phase dynamics corresponding to small
frequencies $\Omega \ll 1$. Due to the Josephson oscillations, the modes with
different frequencies are mixed and the slow mode with frequency $\Omega $ is
coupled to the fast modes with frequencies $\Omega \pm \omega_J $. Near the cavity
resonances the fast modes are resonantly enhanced and one can neglect coupling to
the higher-frequency modes. Therefore the dominant contribution is given by
$(k_{x}\approx 0,k_{ym},\Omega )$ and $(k_{x1},k_{ym},\Omega \pm \omega_J )$.
\cite{Koshelev10} The solution can be written as
\begin{equation}
\tilde{\theta}(\mathrm{r},t)=\sum_{p=-1,0,1}\sum_{m=0}^{\infty }{a_{p}}(x,k_{ym})\cos
(k_{ym}y)\exp [i(\Omega +p\omega_J )t],  \label{eq6}
\end{equation}
with ${a_{0}}(x,k_{ym})\!\approx \!{a_{0}}(k_{ym})$ and ${a_{\pm 1}} (x,k_{ym})\!\approx\!
{a_{\pm 1}}(k_{ym})\cos \left( k_{x1}x\right) $. Substituting Eq. \eqref{eq6} into
Eq. \eqref{eq5} and separating each frequency component, we obtain coupled equations
for the slow and fast components. Excluding the fast components leads to
equation for the slow component
\begin{equation}
(-\eta {\Omega ^{2}}+{i}{\beta _{d}}\Omega +c_{\Omega }^{2}k_{ym}^{2}){
a_{0}(k_{x}=0,k_{ym},\Omega )}=\tilde{J}(\omega_J ),  \label{eq7}
\end{equation}
All parameters of this equation have the regular and resonance contributions. In
particular, the dissipation parameter ${\beta _{d}}$ coincides with the reduced
differential conductivity Eq. (\ref{eq4}). The parameters $c_{\Omega }^{2}$ and $
\eta $ are given by, $c_{\Omega }^2=1+F^2{\rm{Re}}\left[( \omega_J ^{2}+i\beta \omega_J
-{k_{x1}^{2}}) ^{-2}\right]/4$ and
\begin{equation}
\eta =1+\frac{F^{2}}{4}\mathrm{Re}\left[ \frac{k_{x1}^{2}+3\omega_J ^{2}-3i\beta
\omega_J +\beta ^{2}}{\left( k_{x1}^{2}-\omega_J ^{2}+i\beta \omega_J \right) ^{3}}
\right].  \label{eq8}
\end{equation}
$\eta \approx 1$ and $c_{\Omega }^{2}\approx 1$ off the resonance and are enhanced near
the resonance as shown in Fig. \ref{f1} (b). It is important to note that the noise
amplitude is also enhanced near the resonance and is proportional to the total
current, Eq. (\ref{eq3}),
\begin{equation}
\left\langle |\tilde{J}(\omega_J )|^{2}\right\rangle =\frac{{2T\beta }}{{{
L_{x}L_{y}}}}\left( 1+\frac{{F^{2}}}{4}\frac{1}{{{{({\omega_J ^{2}}-k_{x1}^{2})
}^{2}}+{\beta ^{2}}{\omega_J ^{2}}}}\right) =\frac{{2T}}{{{L_{x}L_{y}}}}\frac{
J_{\mathrm{ext}}}{V}.  \label{eq8a}
\end{equation}
The phase diffusion constant $D_{0}$ is given by
$D_{0}=\left\langle |\tilde{J}(\omega_J )|^{2}\right\rangle /\beta
_{d}^{2}$ and can be represented as
\begin{equation}
{D_{0}}=2T\frac{IR_{d}^{2}}{V},  \label{eq9}
\end{equation}
where $I=J_{\mathrm{ext}}L_{x}L_{y}$ is the current and $R_{d}=1/(\beta
_{d}L_{x}L_{y})=dV/dI$ is the differential resistance. It is important to emphasize
that in this nonequilibrium regime the FDT is violated for the slow mode $\langle
\tilde{J}(t)\tilde{J}(t^{\prime })\rangle \neq 2T\beta_d \delta (t-t^{\prime })$.
The spectrum for the $a_{0}$ mode is ${\Omega ^{2}}(k_{ym})=({i}{\beta _{d}}\Omega
+c_{\Omega }^{2}k_{ym}^{2})/\eta $, which becomes gapless when $k_{ym}\rightarrow 0$
as a consequence of the invariance with respect to constant phase shift. Thus this
diffusive mode is most important for the linewidth broadening, and we will
consider this mode in the following calculations of the linewidth.

In the presence of slow fluctuating phase,
\[
\tilde{\theta}_{0}=\int \frac{d\Omega }{2\pi } \sum_{m}a_{0}\cos (k_{ym}y)\exp
(i\Omega t),
\]
the supercurrent density
$ J_{s}(x,y)=g(x)\sin
\left[ \omega _{J}t+\tilde{\theta}_{0}\right] $
is also fluctuating which gives rise to the nonzero linewidth. Here we have neglected
the weak plasma oscillation inside the sine function. The fluctuating plasma
oscillation $ \tilde{\phi}({k_{x1}},{k_{ym}},\omega )$ is given by
\begin{equation}
\tilde{\phi}=\frac{-F\int {d}tdy\sin ({\omega _{J}}t+{\tilde{\theta}_{0}}
)\exp (-{i}\omega t)}{{L_{y}}\left( {k_{x1}^{2}-{\omega _{J}^{2}}+{i}\beta
\omega _{J}}\right) }.  \label{eq10}
\end{equation}
{\color{black}
The linewidth is determined by the spectrum density $S=\langle \tilde{\phi}({
k_{x1}},{0},\omega )\tilde{\phi}({-k_{x1}},{0},-\omega )\rangle $
\begin{equation}
S=\frac{F^{2}\int dtdy\cos (\omega _{J}t)\exp (-i\omega t)\exp
[-K_{-}(y,t)/2]}{(k_{x1}-\omega _{J})^{2}+(\beta \omega _{J})^{2}}.
\label{eq11}
\end{equation}
where $ K_{-} (y,t) = \langle {[{\theta _0}(y,t) - \theta (0,0)]^2}\rangle$ is the
fluctuation phase correlation function which can be approximately evaluated as
\begin{equation}
{K_{-}}(y,t)\approx 2W+{D_{0}}\left[ t-\frac{\eta }{{\beta _{d}}}\left( 1-\exp (-{
{t{\beta _{d}}}}/{\eta })\right) \right].  \label{eq12}
\end{equation}
Here
\[
{W(y,t)}= \frac{{D_0}\beta _d^2}{2}\sum_{m > 0} \int{{d}}\Omega \left(\frac{{1 - \exp [{{i}}({k_{ym}}y + \Omega t)]}}{{{{(\eta {\Omega ^2} - c_\Omega^2k_{ym}^2)}^2} + {{({\beta _d}\Omega )}^2}}}\right),
\]
accounts for the contribution from the gapped modes with $k_{ym}>0$ and the rest
term in ${K_{-}}(y,t)$ accounts for the diffusive mode with $k_{ym}=0$. In the interesting region where $\beta_d\ll D_0$ and $L_y\le 4\pi c_{\Omega}\sqrt{\eta}/\beta_d$,
we
obtain
\begin{equation}
W={{\pi {D_0}{\beta _d}L_y^2}}/({{6c_\Omega^2}}),
\end{equation}
which becomes independent on time and coordinate. $W$ accounts for the suppression
of the radiation and is known as the Debye-Waller factor.} The Debye-Waller factor
decreases near the resonance as shown in Fig. \ref{f1} (d). The broadening of the
linewidth is due to the second term in $K_{-}$. In the region when ${D_{0}}/2\gg
{\beta _{d}}/\eta $,
$K_{-}(t)={D_{0}}{\beta _{d}}{t^{2}}/(2\eta )$ and the line shape is
\begin{equation}
S=\frac{{{F^{2}L_y\exp (-W/2)}}}{{{{
(k_{x1}^{2}-\omega _{J}^{2})}^{2}}+{{(\beta {\omega _{J}})}^{2}}}}\exp \left[
\frac{{\ -{{(\omega -{\omega _{J}})}^{2}}}}{{{D_{0}}{\beta _{d}}/(2\eta )}}
\right] .  \label{eq14}
\end{equation}
The line shape is Gaussian with the linewidth $\Gamma
_{G}\equiv\Delta\omega/(2\pi)={D_{0}}{\beta _{d} }/(8\pi\eta )$. In the other limit
${D_{0}}/2\ll {\beta _{d}}/\eta $, the line shape is determined by slow phase diffusion at large times $K_{-}(t)\approx D_{0}t$
\begin{equation}
S=\frac{D_{0}{{F^{2}}L_y\exp (-{D_{0}}{\eta }/{\beta _{d}})}\exp (-W/2)}{{2[{{(k_{x1}^{2}-\omega _{J}^{2})}^{2}}+{{(\beta {
\omega _{J}})}^{2}}][{{{{(\omega -{\omega _{J}})}^{2}}+D_{0}^{2}/4}}}]}.
\label{eq15}
\end{equation}
The line has Lorentzian shape with the width $\Gamma _{L}=D_{0}/(2\pi)$. 
The diffusion of the gapless perturbation $a_{0}(k_{ym}=0)$ with a diffusion
constant $D_{0}$ caused by thermal fluctuations is responsible for the linewidth
broadening. For typical parameters of Nb-$\mathrm{Al/AlO}_x$-Nb JJs with $L_x=L_y=10\lambda_J$ at low temperature $4.2$ K, we have $\beta\approx 0.1$, $T\approx 10^{-3}$, where $\lambda_J$ is the Josephson length $\lambda_J\approx 10\ \mathrm{\mu m}$. \cite{BaroneBook} The line shape in this region is Lorentzian because $D_{0}/2\ll
\beta _{d}/\eta $. Approaching the
resonance, the linewidth decreases significantly as shown in Fig. \ref{f1} (c). In
both cases, the line shape is proportional to $T$ and is inversely proportional to
the lateral area of the junction $ L_{x}L_{y}$. This behavior is very natural.
Qualitatively, one may treat the JJ as a two-dimensional ensemble of oscillators. If these oscillators are synchronized, the
linewidth is sharpened as the inverse of the population of oscillators, which
 is proportional to the junction area $L_{x}L_{y}$. \cite{Jain84} However as temperature is increased to close to $T_{c}$, the line
shape evolves into Gaussian. The crossover from Lorentzian to
Gaussian line shape occurs at the temperature $T^{\ast }=\beta _{d}^{3}\omega_J
L_{x}L_{y}/(\eta J_{\mathrm{ext}})$. Such Lorentzian-to-Gaussian crossover in the 
line shape was predicted for a point junction \cite{Larkin1968} and for the
Josephson flux flow region \cite{Pankratov2002}.

\begin{figure}[t]
\psfig{figure=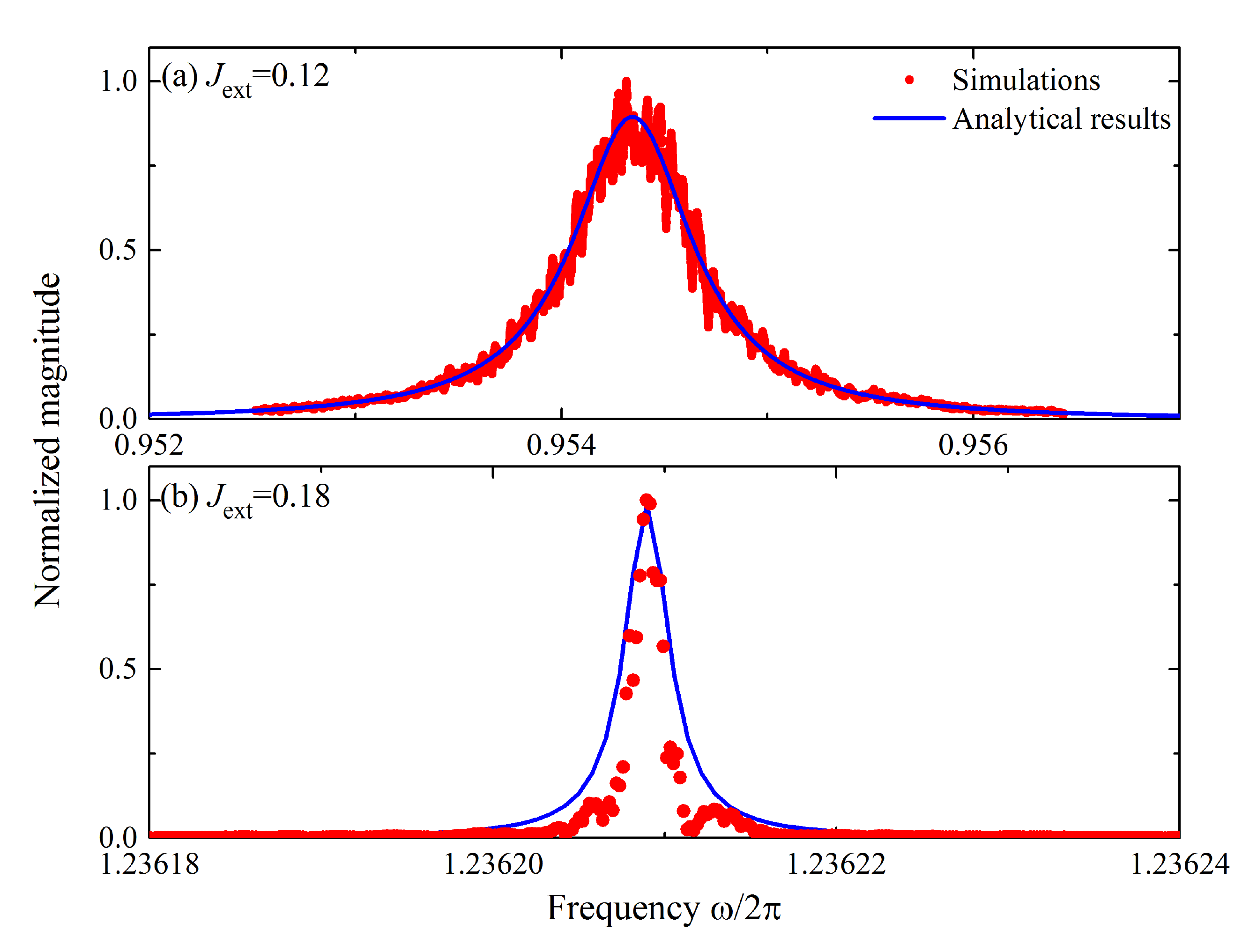,width=\columnwidth}
\caption{(color online) Comparison of the line shape obtained analytically
and numerically both off the resonance (a) and at the resonance (b). The linewidth
sharpen significantly at the cavity resonance. Parameters are the same as
those in Fig. \protect\ref{f1} and $T=1.778\times 10^{-5}$.
} \label{f2}
\end{figure}

For comparison, we performed numerical calculation of Eq. \eqref{eq1}. We
assumed that the system is uniform along the $y$ direction meaning
that the only $k_{ym}=0$ mode is taken into account. We calculated the ac
electric field at one edge of the JJ at $x=0$ and then performed the Fourier
transform to obtain the spectrum. The results for numerical calculations and
analytical treatment are shown in Fig. \ref{f2} . Off the resonance there is a
perfect agreement between two approaches. When the voltage $V=\omega_J$ is tuned
close to the cavity resonance, the amplitude of the plasma oscillation $A$
increases and our theory based on linear expansion becomes inaccurate in
this strongly nonlinear region. There is a small discrepancy between the
analytical and numerical results.

We next proceed to study the radiation linewidth for a stack of free-standing IJJs.
The dynamics of the phase difference $\theta_l$ in the $ l $-th junction are
described by \cite{Sakai93,Bulaevskii94,Bulaevskii96,Machida99,Koshelev01,Hu10, units}
\begin{align}  \label{eq16}
[(1 + {\beta _{ab}}{\partial _t}) - \zeta {\Delta ^{(2)}}][\sin{\theta _l} +
{\beta _c}{\partial _t}{\theta _l} + \partial _t^2{\theta _l} + {\tilde j_z}
(\mathbf{r},l,t)]  \notag \\
= (1 + {\beta _{ab}}{\partial _t})\nabla _{2d}^2{\theta _l} + \nabla_{2d}
\cdot[{\mathbf{\tilde j}_{ab}}(\mathbf{r},l + 1,t) - {\mathbf{\tilde j}_{ab}}(\mathbf{r},l,t)],
\end{align}
where $\Delta^{(2)}h_l \equiv h_{l+1}+h_{l-1}-2 h_l$ is the finite difference
operator, $\zeta=(\lambda_{ab}/s)^2$ is the inductive coupling, $ \beta_c$ and
$\beta_{ab}$ are dissipation parameters due the out-of-plane and in-plane quasiparticle conductivities. \cite{units} The out-of-plane
$\tilde{j}_z$ and in-plane Gaussian noise current have the correlators
\begin{equation}
\langle {\tilde j_z}(\mathbf{r},l,t){\tilde j_z}(\mathbf{r}^{\prime
},l^{\prime },t^{\prime })\rangle = 2T{\beta _c}\zeta^{-1}\delta_{l l^{\prime
}}\delta (\mathbf{r} - \mathbf{r}^{\prime })\delta (t - t^{\prime }),
\end{equation}
\begin{equation}
\langle {\tilde j_\mu }(\mathbf{r},l,t){\tilde j_\upsilon }(\mathbf{r}
^{\prime },l^{\prime },t^{\prime })\rangle = 2T{\beta _{ab}} \delta_{l l^{\prime
}}\delta (\mathbf{r} - \mathbf{r}^{\prime })\delta (t -
t^{\prime }){\delta _{\mu \upsilon }},
\end{equation}
with $\mu, \upsilon=x, y$. When the Josephson frequency approaches the cavity mode
$(k_{x1}, 0)$, the $\pi$ kink state is stabilized \cite{szlin08b,Koshelev08b}
\begin{equation}  \label{eq18}
{\theta _l} = \omega_J t + {\theta _{s,l}}(x) + A\cos \left(\frac{\pi }{{L_x}}
x\right)\exp ({i}\omega_J t),
\end{equation}
where ${\theta _{s,l}}(x)$ runs abruptly from $0$ to $\pi$ at the center of the
junction, which can be treated as a step function ${ \theta
_{s,l}}(x)=[\mathrm{sign}(x-L_x/2)+1]\pi/2$. \cite{Koshelev08b,szlin09a} In the absence of thermal fluctuations, Eq.
\eqref{eq16} reduces to Eq. \eqref{eq1} with a step modulation
$g(x)=-\mathrm{sign}(x-L_x/2)$ \cite{Koshelev10}. For the same reason, the linewidth is due to
the phase diffusion with the gapless mode with $\mathbf{k }=0$, while the the gapped
modes with $\mathbf{k}\neq 0$ contribute to the Debye-Waller factor. 
The most important difference is that the noise amplitude, Eq. \eqref{eq8a} acquires additional $1/N$ factor because the effective noise current for the slow mode is given by averaging over independent noise currents in all synchronized junctions. Here $N$ is the number of junctions. Correspondingly, the phase diffusion coefficient $D_0$ also becomes $N$ times smaller.
The linewidth in IJJs case is again given by Eqs. \eqref{eq14} and \eqref{eq15} with $D_0\rightarrow D_0/N$, and with a different Debye-Waller factor. For the Lorentzian line shape when
${D_0}/2 \ll {\beta _d}/\eta$, the linewidth in terms of $IV$ curve is $\Gamma_L=T I
R_d^2/(\pi VN^2)$, where $R_d$ and $V$ are the total differential resistance and voltage over the whole stack. As $R_d^2/VN$ corresponds to the contribution from one junction and does not depend on $N$, the linewidth of the stack contains an additional $1/N$ factor in comparison with a single junction, Eq. \eqref{eq9}, due to the in-phase oscillations in different junctions.

The linewidth can be expressed in term of $IV$ characteristics. The same expression is also derived long time ago for a point junction.
\cite{Dahm69} For an ideal case when all junctions in the stack are synchronized, we estimate the intrinsic linewidth of the IJJs for $N=600$ to be $\Gamma=0.1$ MHz at $T=4.2$ K, which is a fundamental limit for the THz generator based on BSCCO. In experiments, the junctions are usually partially synchronized, and the linewidth is larger than that in the ideal case. Furthermore, the linewidth decreases with temperature if $R_d(T)$ drops fast when $T$ increases.

For the mesa structures used in experiments, \cite{Ozyuzer07} there is an additional
dissipation due to the radiation into the base crystal \cite{Koshelev09}. This
dissipation can be described using an effective larger damping coefficient
$\beta^{\prime }$. Off the resonance, the linewidth sharpens due to the radiation
into base crystal because $D_0$ decreases with $ \beta$ according to Eq.
\eqref{eq9}. However near the resonance, the linewidth increases since $D_0$ increases with $\beta$ near the resonance.

Finally we discuss the relation between the derived linewidth and the quality factor of the cavity. The quality factor for the JJs in Eq. \eqref{eq1} is $Q=\omega_{nm}/\beta$, and the corresponding linewidth is $\Gamma_c=2Q/\omega_J$. $\Gamma_c$ is a property of the cavity and is independent on the gain medium (Josephson oscillations). $\Gamma_c$ shall be interpreted as the upper bound for the linewidth, which is realized for the completely unsynchronized Josephson oscillations. In the case of synchronized oscillations as we considered here, the phase-diffusion linewidth is much smaller than $\Gamma_c$.

To summarize, we have studied the linewidth of the high frequency electromagnetic radiation from Josephson junctions and a stack of intrinsic Josephson junctions near cavity resonances. The linewidth is caused by the diffusion of the superconducting phase at the gapless mode with wavenumber $\mathbf{k}=0$. The gapped modes with nonzero wave vectors are responsible for the suppression of the radiation power. The linewidth is Lorentzian in low temperature region and can be calculated directly from the $IV$ characteristics. We also predicted a lower bound for the linewidth of the strong terahertz radiation from $\mathrm{{Bi_{2}Sr_{2}CaCu_{2}O_{8}}}$.

\noindent \textit{Acknowledgements --}The authors thanks H. B. Wang, T. M. Benseman, U.
Welp, and L. N. Bulaevskii for helpful discussions. SZL gratefully
acknowledges funding support from the Office of Naval Research via the
Applied Electrodynamics collaboration. AEK is supported by UChicago Argonne,
LLC, operator of Argonne National Laboratory, a US DOE laboratory, operated
under contract No. DE-AC02-06CH11357. Computer resources for numerical calculations were supported by the Institutional Computing Program in LANL.


\begin{thebibliography}{37}%
\makeatletter
\providecommand \@ifxundefined [1]{%
 \@ifx{#1\undefined}
}%
\providecommand \@ifnum [1]{%
 \ifnum #1\expandafter \@firstoftwo
 \else \expandafter \@secondoftwo
 \fi
}%
\providecommand \@ifx [1]{%
 \ifx #1\expandafter \@firstoftwo
 \else \expandafter \@secondoftwo
 \fi
}%
\providecommand \natexlab [1]{#1}%
\providecommand \enquote  [1]{``#1''}%
\providecommand \bibnamefont  [1]{#1}%
\providecommand \bibfnamefont [1]{#1}%
\providecommand \citenamefont [1]{#1}%
\providecommand \href@noop [0]{\@secondoftwo}%
\providecommand \href [0]{\begingroup \@sanitize@url \@href}%
\providecommand \@href[1]{\@@startlink{#1}\@@href}%
\providecommand \@@href[1]{\endgroup#1\@@endlink}%
\providecommand \@sanitize@url [0]{\catcode `\\12\catcode `\$12\catcode
  `\&12\catcode `\#12\catcode `\^12\catcode `\_12\catcode `\%12\relax}%
\providecommand \@@startlink[1]{}%
\providecommand \@@endlink[0]{}%
\providecommand \url  [0]{\begingroup\@sanitize@url \@url }%
\providecommand \@url [1]{\endgroup\@href {#1}{\urlprefix }}%
\providecommand \urlprefix  [0]{URL }%
\providecommand \Eprint [0]{\href }%
\providecommand \doibase [0]{http://dx.doi.org/}%
\providecommand \selectlanguage [0]{\@gobble}%
\providecommand \bibinfo  [0]{\@secondoftwo}%
\providecommand \bibfield  [0]{\@secondoftwo}%
\providecommand \translation [1]{[#1]}%
\providecommand \BibitemOpen [0]{}%
\providecommand \bibitemStop [0]{}%
\providecommand \bibitemNoStop [0]{.\EOS\space}%
\providecommand \EOS [0]{\spacefactor3000\relax}%
\providecommand \BibitemShut  [1]{\csname bibitem#1\endcsname}%
\let\auto@bib@innerbib\@empty
\bibitem [{\citenamefont {Jain}\ \emph {et~al.}(1984)\citenamefont {Jain},
  \citenamefont {Likharev}, \citenamefont {Lukens},\ and\ \citenamefont
  {Sauvageau}}]{Jain84}%
  \BibitemOpen
  \bibfield  {author} {\bibinfo {author} {\bibfnamefont {A.~K.}\ \bibnamefont
  {Jain}}, \bibinfo {author} {\bibfnamefont {K.~K.}\ \bibnamefont {Likharev}},
  \bibinfo {author} {\bibfnamefont {J.~E.}\ \bibnamefont {Lukens}}, \ and\
  \bibinfo {author} {\bibfnamefont {J.~E.}\ \bibnamefont {Sauvageau}},\
  }\href@noop {} {\bibfield  {journal} {\bibinfo  {journal} {Phys. Rep.}\
  }\textbf {\bibinfo {volume} {109}},\ \bibinfo {pages} {309} (\bibinfo {year}
  {1984})}\BibitemShut {NoStop}%
\bibitem [{\citenamefont {Barbara}\ \emph {et~al.}(1999)\citenamefont
  {Barbara}, \citenamefont {Cawthorne}, \citenamefont {Shitov},\ and\
  \citenamefont {Lobb}}]{Barbara99}%
  \BibitemOpen
  \bibfield  {author} {\bibinfo {author} {\bibfnamefont {P.}~\bibnamefont
  {Barbara}}, \bibinfo {author} {\bibfnamefont {A.~B.}\ \bibnamefont
  {Cawthorne}}, \bibinfo {author} {\bibfnamefont {S.~V.}\ \bibnamefont
  {Shitov}}, \ and\ \bibinfo {author} {\bibfnamefont {C.~J.}\ \bibnamefont
  {Lobb}},\ }\href@noop {} {\bibfield  {journal} {\bibinfo  {journal} {Phys.
  Rev. Lett.}\ }\textbf {\bibinfo {volume} {82}},\ \bibinfo {pages} {1963}
  (\bibinfo {year} {1999})}\BibitemShut {NoStop}%
\bibitem [{\citenamefont {Kleiner}\ \emph {et~al.}(1992)\citenamefont
  {Kleiner}, \citenamefont {Steinmeyer}, \citenamefont {Kunkel},\ and\
  \citenamefont {M\"{u}ller}}]{Kleiner92}%
  \BibitemOpen
  \bibfield  {author} {\bibinfo {author} {\bibfnamefont {R.}~\bibnamefont
  {Kleiner}}, \bibinfo {author} {\bibfnamefont {F.}~\bibnamefont {Steinmeyer}},
  \bibinfo {author} {\bibfnamefont {G.}~\bibnamefont {Kunkel}}, \ and\ \bibinfo
  {author} {\bibfnamefont {P.}~\bibnamefont {M\"{u}ller}},\ }\href@noop {}
  {\bibfield  {journal} {\bibinfo  {journal} {Phys. Rev. Lett.}\ }\textbf
  {\bibinfo {volume} {68}},\ \bibinfo {pages} {2394} (\bibinfo {year}
  {1992})}\BibitemShut {NoStop}%
\bibitem [{\citenamefont {Ozyuzer}\ \emph {et~al.}(2007)\citenamefont
  {Ozyuzer}, \citenamefont {Koshelev}, \citenamefont {Kurter}, \citenamefont
  {Gopalsami}, \citenamefont {Li}, \citenamefont {Tachiki}, \citenamefont
  {Kadowaki}, \citenamefont {Yamamoto}, \citenamefont {Minami}, \citenamefont
  {Yamaguchi}, \citenamefont {Tachiki}, \citenamefont {Gray}, \citenamefont
  {Kwok},\ and\ \citenamefont {Welp}}]{Ozyuzer07}%
  \BibitemOpen
  \bibfield  {author} {\bibinfo {author} {\bibfnamefont {L.}~\bibnamefont
  {Ozyuzer}}, \bibinfo {author} {\bibfnamefont {A.~E.}\ \bibnamefont
  {Koshelev}}, \bibinfo {author} {\bibfnamefont {C.}~\bibnamefont {Kurter}},
  \bibinfo {author} {\bibfnamefont {N.}~\bibnamefont {Gopalsami}}, \bibinfo
  {author} {\bibfnamefont {Q.}~\bibnamefont {Li}}, \bibinfo {author}
  {\bibfnamefont {M.}~\bibnamefont {Tachiki}}, \bibinfo {author} {\bibfnamefont
  {K.}~\bibnamefont {Kadowaki}}, \bibinfo {author} {\bibfnamefont
  {T.}~\bibnamefont {Yamamoto}}, \bibinfo {author} {\bibfnamefont
  {H.}~\bibnamefont {Minami}}, \bibinfo {author} {\bibfnamefont
  {H.}~\bibnamefont {Yamaguchi}}, \bibinfo {author} {\bibfnamefont
  {T.}~\bibnamefont {Tachiki}}, \bibinfo {author} {\bibfnamefont {K.~E.}\
  \bibnamefont {Gray}}, \bibinfo {author} {\bibfnamefont {W.~K.}\ \bibnamefont
  {Kwok}}, \ and\ \bibinfo {author} {\bibfnamefont {U.}~\bibnamefont {Welp}},\
  }\href@noop {} {\bibfield  {journal} {\bibinfo  {journal} {Science}\ }\textbf
  {\bibinfo {volume} {318}},\ \bibinfo {pages} {1291} (\bibinfo {year}
  {2007})}\BibitemShut {NoStop}%
\bibitem [{\citenamefont {Bulaevskii}\ and\ \citenamefont
  {Koshelev}(2007)}]{Bulaevskii07}%
  \BibitemOpen
  \bibfield  {author} {\bibinfo {author} {\bibfnamefont {L.~N.}\ \bibnamefont
  {Bulaevskii}}\ and\ \bibinfo {author} {\bibfnamefont {A.~E.}\ \bibnamefont
  {Koshelev}},\ }\href@noop {} {\bibfield  {journal} {\bibinfo  {journal}
  {Phys. Rev. Lett.}\ }\textbf {\bibinfo {volume} {99}},\ \bibinfo {pages}
  {057002} (\bibinfo {year} {2007})}\BibitemShut {NoStop}%
\bibitem [{\citenamefont {Kadowaki}\ \emph {et~al.}(2008)\citenamefont
  {Kadowaki}, \citenamefont {Yamaguchi}, \citenamefont {Kawamata},
  \citenamefont {Yamamoto}, \citenamefont {Minami}, \citenamefont {Kakeya},
  \citenamefont {Welp}, \citenamefont {Ozyuzer}, \citenamefont {Koshelev},
  \citenamefont {Kurter}, \citenamefont {Gray},\ and\ \citenamefont
  {Kwok}}]{kadowaki08}%
  \BibitemOpen
  \bibfield  {author} {\bibinfo {author} {\bibfnamefont {K.}~\bibnamefont
  {Kadowaki}}, \bibinfo {author} {\bibfnamefont {H.}~\bibnamefont {Yamaguchi}},
  \bibinfo {author} {\bibfnamefont {K.}~\bibnamefont {Kawamata}}, \bibinfo
  {author} {\bibfnamefont {T.}~\bibnamefont {Yamamoto}}, \bibinfo {author}
  {\bibfnamefont {H.}~\bibnamefont {Minami}}, \bibinfo {author} {\bibfnamefont
  {I.}~\bibnamefont {Kakeya}}, \bibinfo {author} {\bibfnamefont
  {U.}~\bibnamefont {Welp}}, \bibinfo {author} {\bibfnamefont {L.}~\bibnamefont
  {Ozyuzer}}, \bibinfo {author} {\bibfnamefont {A.}~\bibnamefont {Koshelev}},
  \bibinfo {author} {\bibfnamefont {C.}~\bibnamefont {Kurter}}, \bibinfo
  {author} {\bibfnamefont {K.}~\bibnamefont {Gray}}, \ and\ \bibinfo {author}
  {\bibfnamefont {W.-K.}\ \bibnamefont {Kwok}},\ }\href@noop {} {\bibfield
  {journal} {\bibinfo  {journal} {Physca C}\ }\textbf {\bibinfo {volume}
  {468}},\ \bibinfo {pages} {634} (\bibinfo {year} {2008})}\BibitemShut
  {NoStop}%
\bibitem [{\citenamefont {Wang}\ \emph {et~al.}(2009)\citenamefont {Wang},
  \citenamefont {Gu\'{e}non}, \citenamefont {Yuan}, \citenamefont {Iishi},
  \citenamefont {Arisawa}, \citenamefont {Hatano}, \citenamefont {Yamashita},
  \citenamefont {Koelle},\ and\ \citenamefont {Kleiner}}]{Wang09}%
  \BibitemOpen
  \bibfield  {author} {\bibinfo {author} {\bibfnamefont {H.~B.}\ \bibnamefont
  {Wang}}, \bibinfo {author} {\bibfnamefont {S.}~\bibnamefont {Gu\'{e}non}},
  \bibinfo {author} {\bibfnamefont {J.}~\bibnamefont {Yuan}}, \bibinfo {author}
  {\bibfnamefont {A.}~\bibnamefont {Iishi}}, \bibinfo {author} {\bibfnamefont
  {S.}~\bibnamefont {Arisawa}}, \bibinfo {author} {\bibfnamefont
  {T.}~\bibnamefont {Hatano}}, \bibinfo {author} {\bibfnamefont
  {T.}~\bibnamefont {Yamashita}}, \bibinfo {author} {\bibfnamefont
  {D.}~\bibnamefont {Koelle}}, \ and\ \bibinfo {author} {\bibfnamefont
  {R.}~\bibnamefont {Kleiner}},\ }\href@noop {} {\bibfield  {journal} {\bibinfo
   {journal} {Phys. Rev. Lett.}\ }\textbf {\bibinfo {volume} {102}},\ \bibinfo
  {pages} {017006} (\bibinfo {year} {2009})}\BibitemShut {NoStop}%
\bibitem [{\citenamefont {Wang}\ \emph {et~al.}(2010)\citenamefont {Wang},
  \citenamefont {Gu\'{e}non}, \citenamefont {Gross}, \citenamefont {Yuan},
  \citenamefont {Jiang}, \citenamefont {Zhong}, \citenamefont {Grunzweig},
  \citenamefont {Iishi}, \citenamefont {Wu}, \citenamefont {Hatano},
  \citenamefont {Koelle},\ and\ \citenamefont {Kleiner}}]{Wang10}%
  \BibitemOpen
  \bibfield  {author} {\bibinfo {author} {\bibfnamefont {H.~B.}\ \bibnamefont
  {Wang}}, \bibinfo {author} {\bibfnamefont {S.}~\bibnamefont {Gu\'{e}non}},
  \bibinfo {author} {\bibfnamefont {B.}~\bibnamefont {Gross}}, \bibinfo
  {author} {\bibfnamefont {J.}~\bibnamefont {Yuan}}, \bibinfo {author}
  {\bibfnamefont {Z.~G.}\ \bibnamefont {Jiang}}, \bibinfo {author}
  {\bibfnamefont {Y.~Y.}\ \bibnamefont {Zhong}}, \bibinfo {author}
  {\bibfnamefont {M.}~\bibnamefont {Grunzweig}}, \bibinfo {author}
  {\bibfnamefont {A.}~\bibnamefont {Iishi}}, \bibinfo {author} {\bibfnamefont
  {P.~H.}\ \bibnamefont {Wu}}, \bibinfo {author} {\bibfnamefont
  {T.}~\bibnamefont {Hatano}}, \bibinfo {author} {\bibfnamefont
  {D.}~\bibnamefont {Koelle}}, \ and\ \bibinfo {author} {\bibfnamefont
  {R.}~\bibnamefont {Kleiner}},\ }\href@noop {} {\bibfield  {journal} {\bibinfo
   {journal} {Phys. Rev. Lett.}\ }\textbf {\bibinfo {volume} {105}},\ \bibinfo
  {pages} {057002} (\bibinfo {year} {2010})}\BibitemShut {NoStop}%
\bibitem [{\citenamefont {Tsujimoto}\ \emph {et~al.}(2010)\citenamefont
  {Tsujimoto}, \citenamefont {Yamaki}, \citenamefont {Deguchi}, \citenamefont
  {Yamamoto}, \citenamefont {Kashiwagi}, \citenamefont {Minami}, \citenamefont
  {Tachiki}, \citenamefont {Kadowaki},\ and\ \citenamefont
  {Klemm}}]{Tsujimoto10}%
  \BibitemOpen
  \bibfield  {author} {\bibinfo {author} {\bibfnamefont {M.}~\bibnamefont
  {Tsujimoto}}, \bibinfo {author} {\bibfnamefont {K.}~\bibnamefont {Yamaki}},
  \bibinfo {author} {\bibfnamefont {K.}~\bibnamefont {Deguchi}}, \bibinfo
  {author} {\bibfnamefont {T.}~\bibnamefont {Yamamoto}}, \bibinfo {author}
  {\bibfnamefont {T.}~\bibnamefont {Kashiwagi}}, \bibinfo {author}
  {\bibfnamefont {H.}~\bibnamefont {Minami}}, \bibinfo {author} {\bibfnamefont
  {M.}~\bibnamefont {Tachiki}}, \bibinfo {author} {\bibfnamefont
  {K.}~\bibnamefont {Kadowaki}}, \ and\ \bibinfo {author} {\bibfnamefont
  {R.~A.}\ \bibnamefont {Klemm}},\ }\href@noop {} {\bibfield  {journal}
  {\bibinfo  {journal} {Phys. Rev. Lett.}\ }\textbf {\bibinfo {volume} {105}},\
  \bibinfo {pages} {037005} (\bibinfo {year} {2010})}\BibitemShut {NoStop}%
\bibitem [{\citenamefont {Kashiwagi}\ \emph {et~al.}(2012)\citenamefont
  {Kashiwagi}, \citenamefont {Tsujimoto}, \citenamefont {Yamamoto},
  \citenamefont {Minami}, \citenamefont {Yamaki}, \citenamefont {Delfanazari},
  \citenamefont {Deguchi}, \citenamefont {Orita}, \citenamefont {Koike},
  \citenamefont {Nakayama}, \citenamefont {Kitamura}, \citenamefont {Sawamura},
  \citenamefont {Hagino}, \citenamefont {Ishida}, \citenamefont {Ivanovic},
  \citenamefont {Asai}, \citenamefont {Tachiki}, \citenamefont {Klemm},\ and\
  \citenamefont {Kadowaki}}]{Kashiwagi2012}%
  \BibitemOpen
  \bibfield  {author} {\bibinfo {author} {\bibfnamefont {T.}~\bibnamefont
  {Kashiwagi}}, \bibinfo {author} {\bibfnamefont {M.}~\bibnamefont
  {Tsujimoto}}, \bibinfo {author} {\bibfnamefont {T.}~\bibnamefont {Yamamoto}},
  \bibinfo {author} {\bibfnamefont {H.}~\bibnamefont {Minami}}, \bibinfo
  {author} {\bibfnamefont {K.}~\bibnamefont {Yamaki}}, \bibinfo {author}
  {\bibfnamefont {K.}~\bibnamefont {Delfanazari}}, \bibinfo {author}
  {\bibfnamefont {K.}~\bibnamefont {Deguchi}}, \bibinfo {author} {\bibfnamefont
  {N.}~\bibnamefont {Orita}}, \bibinfo {author} {\bibfnamefont
  {T.}~\bibnamefont {Koike}}, \bibinfo {author} {\bibfnamefont
  {R.}~\bibnamefont {Nakayama}}, \bibinfo {author} {\bibfnamefont
  {T.}~\bibnamefont {Kitamura}}, \bibinfo {author} {\bibfnamefont
  {M.}~\bibnamefont {Sawamura}}, \bibinfo {author} {\bibfnamefont
  {S.}~\bibnamefont {Hagino}}, \bibinfo {author} {\bibfnamefont
  {K.}~\bibnamefont {Ishida}}, \bibinfo {author} {\bibfnamefont
  {K.}~\bibnamefont {Ivanovic}}, \bibinfo {author} {\bibfnamefont
  {H.}~\bibnamefont {Asai}}, \bibinfo {author} {\bibfnamefont {M.}~\bibnamefont
  {Tachiki}}, \bibinfo {author} {\bibfnamefont {R.~A.}\ \bibnamefont {Klemm}},
  \ and\ \bibinfo {author} {\bibfnamefont {K.}~\bibnamefont {Kadowaki}},\
  }\href {\doibase 10.1143/JJAP.51.010113} {\bibfield  {journal} {\bibinfo
  {journal} {Jpn. J. Appl. Phys.}\ }\textbf {\bibinfo {volume} {51}},\ \bibinfo
  {pages} {010113} (\bibinfo {year} {2012})}\BibitemShut {NoStop}%
\bibitem [{\citenamefont {Tsujimoto}\ \emph {et~al.}(2012)\citenamefont
  {Tsujimoto}, \citenamefont {Yamamoto}, \citenamefont {Delfanazari},
  \citenamefont {Nakayama}, \citenamefont {Kitamura}, \citenamefont {Sawamura},
  \citenamefont {Kashiwagi}, \citenamefont {Minami}, \citenamefont {Tachiki},
  \citenamefont {Kadowaki},\ and\ \citenamefont {Klemm}}]{Tsujimoto12b}%
  \BibitemOpen
  \bibfield  {author} {\bibinfo {author} {\bibfnamefont {M.}~\bibnamefont
  {Tsujimoto}}, \bibinfo {author} {\bibfnamefont {T.}~\bibnamefont {Yamamoto}},
  \bibinfo {author} {\bibfnamefont {K.}~\bibnamefont {Delfanazari}}, \bibinfo
  {author} {\bibfnamefont {R.}~\bibnamefont {Nakayama}}, \bibinfo {author}
  {\bibfnamefont {T.}~\bibnamefont {Kitamura}}, \bibinfo {author}
  {\bibfnamefont {M.}~\bibnamefont {Sawamura}}, \bibinfo {author}
  {\bibfnamefont {T.}~\bibnamefont {Kashiwagi}}, \bibinfo {author}
  {\bibfnamefont {H.}~\bibnamefont {Minami}}, \bibinfo {author} {\bibfnamefont
  {M.}~\bibnamefont {Tachiki}}, \bibinfo {author} {\bibfnamefont
  {K.}~\bibnamefont {Kadowaki}}, \ and\ \bibinfo {author} {\bibfnamefont
  {R.~A.}\ \bibnamefont {Klemm}},\ }\href@noop {} {\bibfield  {journal}
  {\bibinfo  {journal} {Phys. Rev. Lett.}\ }\textbf {\bibinfo {volume} {108}},\
  \bibinfo {pages} {107006} (\bibinfo {year} {2012})}\BibitemShut {NoStop}%
\bibitem [{\citenamefont {Benseman}\ \emph {et~al.}(2011)\citenamefont
  {Benseman}, \citenamefont {Koshelev}, \citenamefont {Gray}, \citenamefont
  {Kwok}, \citenamefont {Welp}, \citenamefont {Kadowaki}, \citenamefont
  {Tachiki},\ and\ \citenamefont {Yamamoto}}]{Benseman11}%
  \BibitemOpen
  \bibfield  {author} {\bibinfo {author} {\bibfnamefont {T.~M.}\ \bibnamefont
  {Benseman}}, \bibinfo {author} {\bibfnamefont {A.~E.}\ \bibnamefont
  {Koshelev}}, \bibinfo {author} {\bibfnamefont {K.~E.}\ \bibnamefont {Gray}},
  \bibinfo {author} {\bibfnamefont {W.-K.}\ \bibnamefont {Kwok}}, \bibinfo
  {author} {\bibfnamefont {U.}~\bibnamefont {Welp}}, \bibinfo {author}
  {\bibfnamefont {K.}~\bibnamefont {Kadowaki}}, \bibinfo {author}
  {\bibfnamefont {M.}~\bibnamefont {Tachiki}}, \ and\ \bibinfo {author}
  {\bibfnamefont {T.}~\bibnamefont {Yamamoto}},\ }\href@noop {} {\bibfield
  {journal} {\bibinfo  {journal} {Phys. Rev. B}\ }\textbf {\bibinfo {volume}
  {84}},\ \bibinfo {pages} {064523} (\bibinfo {year} {2011})}\BibitemShut
  {NoStop}%
\bibitem [{\citenamefont {Orita}\ \emph {et~al.}(2010)\citenamefont {Orita},
  \citenamefont {Minami}, \citenamefont {Koike}, \citenamefont {Yamamoto},\
  and\ \citenamefont {Kadowaki}}]{Orita2010}%
  \BibitemOpen
  \bibfield  {author} {\bibinfo {author} {\bibfnamefont {N.}~\bibnamefont
  {Orita}}, \bibinfo {author} {\bibfnamefont {H.}~\bibnamefont {Minami}},
  \bibinfo {author} {\bibfnamefont {T.}~\bibnamefont {Koike}}, \bibinfo
  {author} {\bibfnamefont {T.}~\bibnamefont {Yamamoto}}, \ and\ \bibinfo
  {author} {\bibfnamefont {K.}~\bibnamefont {Kadowaki}},\ }\href@noop {}
  {\bibfield  {journal} {\bibinfo  {journal} {Physica C}\ }\textbf {\bibinfo
  {volume} {470}},\ \bibinfo {pages} {S786} (\bibinfo {year}
  {2010})}\BibitemShut {NoStop}%
\bibitem [{\citenamefont {Lin}\ \emph {et~al.}(2008)\citenamefont {Lin},
  \citenamefont {Hu},\ and\ \citenamefont {Tachiki}}]{szlin08}%
  \BibitemOpen
  \bibfield  {author} {\bibinfo {author} {\bibfnamefont {S.~Z.}\ \bibnamefont
  {Lin}}, \bibinfo {author} {\bibfnamefont {X.}~\bibnamefont {Hu}}, \ and\
  \bibinfo {author} {\bibfnamefont {M.}~\bibnamefont {Tachiki}},\ }\href@noop
  {} {\bibfield  {journal} {\bibinfo  {journal} {Phys. Rev.}\ }\textbf
  {\bibinfo {volume} {B77}},\ \bibinfo {pages} {014507} (\bibinfo {year}
  {2008})}\BibitemShut {NoStop}%
\bibitem [{\citenamefont {Lin}\ and\ \citenamefont {Hu}(2008)}]{szlin08b}%
  \BibitemOpen
  \bibfield  {author} {\bibinfo {author} {\bibfnamefont {S.~Z.}\ \bibnamefont
  {Lin}}\ and\ \bibinfo {author} {\bibfnamefont {X.}~\bibnamefont {Hu}},\
  }\href@noop {} {\bibfield  {journal} {\bibinfo  {journal} {Phys. Rev. Lett.}\
  }\textbf {\bibinfo {volume} {100}},\ \bibinfo {pages} {247006} (\bibinfo
  {year} {2008})}\BibitemShut {NoStop}%
\bibitem [{\citenamefont {Koshelev}(2008)}]{Koshelev08b}%
  \BibitemOpen
  \bibfield  {author} {\bibinfo {author} {\bibfnamefont {A.~E.}\ \bibnamefont
  {Koshelev}},\ }\href@noop {} {\bibfield  {journal} {\bibinfo  {journal}
  {Phys. Rev. B}\ }\textbf {\bibinfo {volume} {78}},\ \bibinfo {pages} {174509}
  (\bibinfo {year} {2008})}\BibitemShut {NoStop}%
\bibitem [{\citenamefont {Lin}\ and\ \citenamefont {Hu}(2012)}]{szlin12a}%
  \BibitemOpen
  \bibfield  {author} {\bibinfo {author} {\bibfnamefont {S.-Z.}\ \bibnamefont
  {Lin}}\ and\ \bibinfo {author} {\bibfnamefont {X.}~\bibnamefont {Hu}},\
  }\href {\doibase 10.1103/PhysRevB.86.054506} {\bibfield  {journal} {\bibinfo
  {journal} {Phys. Rev. B}\ }\textbf {\bibinfo {volume} {86}},\ \bibinfo
  {pages} {054506} (\bibinfo {year} {2012})}\BibitemShut {NoStop}%
\bibitem [{\citenamefont {An}\ \emph {et~al.}(2013)\citenamefont {An},
  \citenamefont {Yuan}, \citenamefont {Kinev}, \citenamefont {Li},
  \citenamefont {Huang}, \citenamefont {Ji}, \citenamefont {Zhang},
  \citenamefont {Sun}, \citenamefont {Kang}, \citenamefont {Jin}, \citenamefont
  {Chen}, \citenamefont {Li}, \citenamefont {Gross}, \citenamefont {Ishii},
  \citenamefont {Hirata}, \citenamefont {Hatano}, \citenamefont {Koshelets},
  \citenamefont {Koelle}, \citenamefont {Kleiner}, \citenamefont {Wang},
  \citenamefont {Xu},\ and\ \citenamefont {Wu}}]{An2013}%
  \BibitemOpen
  \bibfield  {author} {\bibinfo {author} {\bibfnamefont {D.~Y.}\ \bibnamefont
  {An}}, \bibinfo {author} {\bibfnamefont {J.}~\bibnamefont {Yuan}}, \bibinfo
  {author} {\bibfnamefont {N.}~\bibnamefont {Kinev}}, \bibinfo {author}
  {\bibfnamefont {M.~Y.}\ \bibnamefont {Li}}, \bibinfo {author} {\bibfnamefont
  {Y.}~\bibnamefont {Huang}}, \bibinfo {author} {\bibfnamefont
  {M.}~\bibnamefont {Ji}}, \bibinfo {author} {\bibfnamefont {H.}~\bibnamefont
  {Zhang}}, \bibinfo {author} {\bibfnamefont {Z.~L.}\ \bibnamefont {Sun}},
  \bibinfo {author} {\bibfnamefont {L.}~\bibnamefont {Kang}}, \bibinfo {author}
  {\bibfnamefont {B.~B.}\ \bibnamefont {Jin}}, \bibinfo {author} {\bibfnamefont
  {J.}~\bibnamefont {Chen}}, \bibinfo {author} {\bibfnamefont {J.}~\bibnamefont
  {Li}}, \bibinfo {author} {\bibfnamefont {B.}~\bibnamefont {Gross}}, \bibinfo
  {author} {\bibfnamefont {A.}~\bibnamefont {Ishii}}, \bibinfo {author}
  {\bibfnamefont {K.}~\bibnamefont {Hirata}}, \bibinfo {author} {\bibfnamefont
  {T.}~\bibnamefont {Hatano}}, \bibinfo {author} {\bibfnamefont {V.~P.}\
  \bibnamefont {Koshelets}}, \bibinfo {author} {\bibfnamefont {D.}~\bibnamefont
  {Koelle}}, \bibinfo {author} {\bibfnamefont {R.}~\bibnamefont {Kleiner}},
  \bibinfo {author} {\bibfnamefont {H.~B.}\ \bibnamefont {Wang}}, \bibinfo
  {author} {\bibfnamefont {W.~W.}\ \bibnamefont {Xu}}, \ and\ \bibinfo {author}
  {\bibfnamefont {P.~H.}\ \bibnamefont {Wu}},\ }\href {\doibase
  doi:10.1063/1.4794072} {\bibfield  {journal} {\bibinfo  {journal} {Appl.
  Phys. Lett.}\ }\textbf {\bibinfo {volume} {102}},\ \bibinfo {pages} {092601}
  (\bibinfo {year} {2013})}\BibitemShut {NoStop}%
\bibitem [{\citenamefont {Tonouchi}(2007)}]{Tonouchi07}%
  \BibitemOpen
  \bibfield  {author} {\bibinfo {author} {\bibfnamefont {M.}~\bibnamefont
  {Tonouchi}},\ }\href@noop {} {\bibfield  {journal} {\bibinfo  {journal} {Nat.
  Photon.}\ }\textbf {\bibinfo {volume} {1}},\ \bibinfo {pages} {97} (\bibinfo
  {year} {2007})}\BibitemShut {NoStop}%
\bibitem [{\citenamefont {Larkin}\ and\ \citenamefont
  {Ovchinnikov}(1968)}]{Larkin1968}%
  \BibitemOpen
  \bibfield  {author} {\bibinfo {author} {\bibfnamefont {A.~I.}\ \bibnamefont
  {Larkin}}\ and\ \bibinfo {author} {\bibfnamefont {Y.~N.}\ \bibnamefont
  {Ovchinnikov}},\ }\href@noop {} {\bibfield  {journal} {\bibinfo  {journal}
  {Sov. Phys. JETP}\ }\textbf {\bibinfo {volume} {26}},\ \bibinfo {pages}
  {1219} (\bibinfo {year} {1968})}\BibitemShut {NoStop}%
\bibitem [{\citenamefont {Stephen}(1968)}]{Stephen68}%
  \BibitemOpen
  \bibfield  {author} {\bibinfo {author} {\bibfnamefont {M.~J.}\ \bibnamefont
  {Stephen}},\ }\href {\doibase 10.1103/PhysRevLett.21.1629} {\bibfield
  {journal} {\bibinfo  {journal} {Phys. Rev. Lett.}\ }\textbf {\bibinfo
  {volume} {21}},\ \bibinfo {pages} {1629} (\bibinfo {year}
  {1968})}\BibitemShut {NoStop}%
\bibitem [{\citenamefont {Dahm}\ \emph {et~al.}(1969)\citenamefont {Dahm},
  \citenamefont {Denenstein}, \citenamefont {Langenberg}, \citenamefont
  {Parker}, \citenamefont {Rogovin},\ and\ \citenamefont {Scalapino}}]{Dahm69}%
  \BibitemOpen
  \bibfield  {author} {\bibinfo {author} {\bibfnamefont {A.~J.}\ \bibnamefont
  {Dahm}}, \bibinfo {author} {\bibfnamefont {A.}~\bibnamefont {Denenstein}},
  \bibinfo {author} {\bibfnamefont {D.~N.}\ \bibnamefont {Langenberg}},
  \bibinfo {author} {\bibfnamefont {W.~H.}\ \bibnamefont {Parker}}, \bibinfo
  {author} {\bibfnamefont {D.}~\bibnamefont {Rogovin}}, \ and\ \bibinfo
  {author} {\bibfnamefont {D.~J.}\ \bibnamefont {Scalapino}},\ }\href {\doibase
  10.1103/PhysRevLett.22.1416} {\bibfield  {journal} {\bibinfo  {journal}
  {Phys. Rev. Lett.}\ }\textbf {\bibinfo {volume} {22}},\ \bibinfo {pages}
  {1416} (\bibinfo {year} {1969})}\BibitemShut {NoStop}%
\bibitem [{\citenamefont {Bulaevskii}\ \emph {et~al.}(2011)\citenamefont
  {Bulaevskii}, \citenamefont {Martin},\ and\ \citenamefont
  {Hal\'asz}}]{Bulaevskii11}%
  \BibitemOpen
  \bibfield  {author} {\bibinfo {author} {\bibfnamefont {L.~N.}\ \bibnamefont
  {Bulaevskii}}, \bibinfo {author} {\bibfnamefont {I.}~\bibnamefont {Martin}},
  \ and\ \bibinfo {author} {\bibfnamefont {G.~B.}\ \bibnamefont {Hal\'asz}},\
  }\href {\doibase 10.1103/PhysRevB.84.014516} {\bibfield  {journal} {\bibinfo
  {journal} {Phys. Rev. B}\ }\textbf {\bibinfo {volume} {84}},\ \bibinfo
  {pages} {014516} (\bibinfo {year} {2011})}\BibitemShut {NoStop}%
\bibitem [{\citenamefont {Li}\ \emph {et~al.}(2012)\citenamefont {Li},
  \citenamefont {Yuan}, \citenamefont {Kinev}, \citenamefont {Li},
  \citenamefont {Gross}, \citenamefont {Gu\'enon}, \citenamefont {Ishii},
  \citenamefont {Hirata}, \citenamefont {Hatano}, \citenamefont {Koelle},
  \citenamefont {Kleiner}, \citenamefont {Koshelets}, \citenamefont {Wang},\
  and\ \citenamefont {Wu}}]{LiLinewidth12}%
  \BibitemOpen
  \bibfield  {author} {\bibinfo {author} {\bibfnamefont {M.}~\bibnamefont
  {Li}}, \bibinfo {author} {\bibfnamefont {J.}~\bibnamefont {Yuan}}, \bibinfo
  {author} {\bibfnamefont {N.}~\bibnamefont {Kinev}}, \bibinfo {author}
  {\bibfnamefont {J.}~\bibnamefont {Li}}, \bibinfo {author} {\bibfnamefont
  {B.}~\bibnamefont {Gross}}, \bibinfo {author} {\bibfnamefont
  {S.}~\bibnamefont {Gu\'enon}}, \bibinfo {author} {\bibfnamefont
  {A.}~\bibnamefont {Ishii}}, \bibinfo {author} {\bibfnamefont
  {K.}~\bibnamefont {Hirata}}, \bibinfo {author} {\bibfnamefont
  {T.}~\bibnamefont {Hatano}}, \bibinfo {author} {\bibfnamefont
  {D.}~\bibnamefont {Koelle}}, \bibinfo {author} {\bibfnamefont
  {R.}~\bibnamefont {Kleiner}}, \bibinfo {author} {\bibfnamefont {V.~P.}\
  \bibnamefont {Koshelets}}, \bibinfo {author} {\bibfnamefont {H.}~\bibnamefont
  {Wang}}, \ and\ \bibinfo {author} {\bibfnamefont {P.}~\bibnamefont {Wu}},\
  }\href {\doibase 10.1103/PhysRevB.86.060505} {\bibfield  {journal} {\bibinfo
  {journal} {Phys. Rev. B}\ }\textbf {\bibinfo {volume} {86}},\ \bibinfo
  {pages} {060505} (\bibinfo {year} {2012})}\BibitemShut {NoStop}%
\bibitem [{\citenamefont {Koshelev}\ and\ \citenamefont
  {Bulaevskii}(2008)}]{Koshelev08}%
  \BibitemOpen
  \bibfield  {author} {\bibinfo {author} {\bibfnamefont {A.~E.}\ \bibnamefont
  {Koshelev}}\ and\ \bibinfo {author} {\bibfnamefont {L.~N.}\ \bibnamefont
  {Bulaevskii}},\ }\href@noop {} {\bibfield  {journal} {\bibinfo  {journal}
  {Phys. Rev. B}\ }\textbf {\bibinfo {volume} {77}},\ \bibinfo {pages} {014530}
  (\bibinfo {year} {2008})}\BibitemShut {NoStop}%
\bibitem [{\citenamefont {Koshelev}(2010)}]{Koshelev10}%
  \BibitemOpen
  \bibfield  {author} {\bibinfo {author} {\bibfnamefont {A.~E.}\ \bibnamefont
  {Koshelev}},\ }\href {\doibase 10.1103/PhysRevB.82.174512} {\bibfield
  {journal} {\bibinfo  {journal} {Phys. Rev. B}\ }\textbf {\bibinfo {volume}
  {82}},\ \bibinfo {pages} {174512} (\bibinfo {year} {2010})}\BibitemShut
  {NoStop}%
\bibitem [{\citenamefont {Barone}\ and\ \citenamefont
  {Paterno}(1982)}]{BaroneBook}%
  \BibitemOpen
  \bibfield  {author} {\bibinfo {author} {\bibfnamefont {A.}~\bibnamefont
  {Barone}}\ and\ \bibinfo {author} {\bibfnamefont {G.}~\bibnamefont
  {Paterno}},\ }\href@noop {} {\emph {\bibinfo {title} {Physics and
  Applications of The Josephson Effect}}}\ (\bibinfo  {publisher} {Wiley},\
  \bibinfo {year} {1982})\BibitemShut {NoStop}%
\bibitem [{\citenamefont {Pankratov}(2002)}]{Pankratov2002}%
  \BibitemOpen
  \bibfield  {author} {\bibinfo {author} {\bibfnamefont {A.~L.}\ \bibnamefont
  {Pankratov}},\ }\href {\doibase 10.1103/PhysRevB.65.054504} {\bibfield
  {journal} {\bibinfo  {journal} {Phys. Rev. B}\ }\textbf {\bibinfo {volume}
  {65}},\ \bibinfo {pages} {054504} (\bibinfo {year} {2002})}\BibitemShut
  {NoStop}%
\bibitem [{\citenamefont {Sakai}\ \emph {et~al.}(1993)\citenamefont {Sakai},
  \citenamefont {Bodin},\ and\ \citenamefont {Pedersen}}]{Sakai93}%
  \BibitemOpen
  \bibfield  {author} {\bibinfo {author} {\bibfnamefont {S.}~\bibnamefont
  {Sakai}}, \bibinfo {author} {\bibfnamefont {P.}~\bibnamefont {Bodin}}, \ and\
  \bibinfo {author} {\bibfnamefont {N.~F.}\ \bibnamefont {Pedersen}},\
  }\href@noop {} {\bibfield  {journal} {\bibinfo  {journal} {J. Appl. Phys.}\
  }\textbf {\bibinfo {volume} {73}},\ \bibinfo {pages} {2411} (\bibinfo {year}
  {1993})}\BibitemShut {NoStop}%
\bibitem [{\citenamefont {Bulaevskii}\ \emph {et~al.}(1994)\citenamefont
  {Bulaevskii}, \citenamefont {Zamora}, \citenamefont {Baeriswyl},
  \citenamefont {Beck},\ and\ \citenamefont {Clem}}]{Bulaevskii94}%
  \BibitemOpen
  \bibfield  {author} {\bibinfo {author} {\bibfnamefont {L.~N.}\ \bibnamefont
  {Bulaevskii}}, \bibinfo {author} {\bibfnamefont {M.}~\bibnamefont {Zamora}},
  \bibinfo {author} {\bibfnamefont {D.}~\bibnamefont {Baeriswyl}}, \bibinfo
  {author} {\bibfnamefont {H.}~\bibnamefont {Beck}}, \ and\ \bibinfo {author}
  {\bibfnamefont {J.~R.}\ \bibnamefont {Clem}},\ }\href@noop {} {\bibfield
  {journal} {\bibinfo  {journal} {Phys. Rev. B}\ }\textbf {\bibinfo {volume}
  {50}},\ \bibinfo {pages} {12831} (\bibinfo {year} {1994})}\BibitemShut
  {NoStop}%
\bibitem [{\citenamefont {Bulaevskii}\ \emph {et~al.}(1996)\citenamefont
  {Bulaevskii}, \citenamefont {Dom\'{\i}nguez}, \citenamefont {Maley},
  \citenamefont {Bishop},\ and\ \citenamefont {Ivlev}}]{Bulaevskii96}%
  \BibitemOpen
  \bibfield  {author} {\bibinfo {author} {\bibfnamefont {L.~N.}\ \bibnamefont
  {Bulaevskii}}, \bibinfo {author} {\bibfnamefont {D.}~\bibnamefont
  {Dom\'{\i}nguez}}, \bibinfo {author} {\bibfnamefont {M.~P.}\ \bibnamefont
  {Maley}}, \bibinfo {author} {\bibfnamefont {A.~R.}\ \bibnamefont {Bishop}}, \
  and\ \bibinfo {author} {\bibfnamefont {B.~I.}\ \bibnamefont {Ivlev}},\
  }\href@noop {} {\bibfield  {journal} {\bibinfo  {journal} {Phys. Rev. B}\
  }\textbf {\bibinfo {volume} {53}},\ \bibinfo {pages} {14601} (\bibinfo {year}
  {1996})}\BibitemShut {NoStop}%
\bibitem [{\citenamefont {Machida}\ \emph {et~al.}(1999)\citenamefont
  {Machida}, \citenamefont {Koyama},\ and\ \citenamefont
  {Tachiki}}]{Machida99}%
  \BibitemOpen
  \bibfield  {author} {\bibinfo {author} {\bibfnamefont {M.}~\bibnamefont
  {Machida}}, \bibinfo {author} {\bibfnamefont {T.}~\bibnamefont {Koyama}}, \
  and\ \bibinfo {author} {\bibfnamefont {M.}~\bibnamefont {Tachiki}},\
  }\href@noop {} {\bibfield  {journal} {\bibinfo  {journal} {Phys. Rev. Lett.}\
  }\textbf {\bibinfo {volume} {83}},\ \bibinfo {pages} {4618} (\bibinfo {year}
  {1999})}\BibitemShut {NoStop}%
\bibitem [{\citenamefont {Koshelev}\ and\ \citenamefont
  {Aranson}(2001)}]{Koshelev01}%
  \BibitemOpen
  \bibfield  {author} {\bibinfo {author} {\bibfnamefont {A.~E.}\ \bibnamefont
  {Koshelev}}\ and\ \bibinfo {author} {\bibfnamefont {I.}~\bibnamefont
  {Aranson}},\ }\href@noop {} {\bibfield  {journal} {\bibinfo  {journal} {Phys.
  Rev. B}\ }\textbf {\bibinfo {volume} {64}},\ \bibinfo {pages} {174508}
  (\bibinfo {year} {2001})}\BibitemShut {NoStop}%
\bibitem [{\citenamefont {Hu}\ and\ \citenamefont {Lin}(2010)}]{Hu10}%
  \BibitemOpen
  \bibfield  {author} {\bibinfo {author} {\bibfnamefont {X.}~\bibnamefont
  {Hu}}\ and\ \bibinfo {author} {\bibfnamefont {S.~Z.}\ \bibnamefont {Lin}},\
  }\href@noop {} {\bibfield  {journal} {\bibinfo  {journal} {Supercond. Sci.
  Technol.}\ }\textbf {\bibinfo {volume} {23}},\ \bibinfo {pages} {053001}
  (\bibinfo {year} {2010})}\BibitemShut {NoStop}%
\bibitem [{uni()}]{units}%
  \BibitemOpen
  \href@noop {} {}\bibinfo {note} {In Eq. (\ref{eq16}), the inductive coupling
  is denfined as $\zeta=(\lambda_{ab}/s)^2$ and the renormalized conductivities
  along the $c$ axis and $ab$ plane are defined as
  $\beta_c=4\pi\sigma_c/(\epsilon_c \omega_p)$ and $\beta_{ab}=4\pi
  \sigma_{ab}\lambda_{ab}^2/(\lambda_c^2\epsilon_c\omega_p)$ with the Josephson
  plasma frequency $\omega_p=c/(\lambda_c\sqrt{\epsilon_c})$. Here
  $\lambda_{ab}$ and $\lambda_c$ are the London penetration depths and $s$ is
  the period of the stack of IJJs. Frequency is in units of $\omega_p$; length
  is in units of $\lambda_c$; current is in units of the Josephson critical
  current density $J_c$; temperature is in units of $\Phi_0^2 s/(16\pi^3
  \lambda_{ab}^2 k_B)$ and magnetic field is in units of $\Phi_0/(2\pi\lambda_c
  s)$ with $\Phi_0=hc/(2e)$ the flux quantum. For BSCCO,
  $\omega_J/(2\pi)\approx 0.1 \rm{\ THz}$, $\lambda_{ab}\approx 0.4\rm{\ \mu
  m}$ and $\lambda_c\approx 200 \rm {\ \mu m}$. The dimensionless electric
  field is given by $E_{z,l}=\partial_t\varphi_l$, with $E$ in units of
  $\Phi_0\omega_p/(2\pi c s)$.}\BibitemShut {Stop}%
\bibitem [{\citenamefont {Lin}\ and\ \citenamefont {Hu}(2009)}]{szlin09a}%
  \BibitemOpen
  \bibfield  {author} {\bibinfo {author} {\bibfnamefont {S.~Z.}\ \bibnamefont
  {Lin}}\ and\ \bibinfo {author} {\bibfnamefont {X.}~\bibnamefont {Hu}},\
  }\href@noop {} {\bibfield  {journal} {\bibinfo  {journal} {Phys. Rev. B}\
  }\textbf {\bibinfo {volume} {79}},\ \bibinfo {pages} {104507} (\bibinfo
  {year} {2009})}\BibitemShut {NoStop}%
\bibitem [{\citenamefont {Koshelev}\ and\ \citenamefont
  {Bulaevskii}(2009)}]{Koshelev09}%
  \BibitemOpen
  \bibfield  {author} {\bibinfo {author} {\bibfnamefont {A.~E.}\ \bibnamefont
  {Koshelev}}\ and\ \bibinfo {author} {\bibfnamefont {L.~N.}\ \bibnamefont
  {Bulaevskii}},\ }\href@noop {} {\bibfield  {journal} {\bibinfo  {journal} {J.
  Phys.: Conference Series}\ }\textbf {\bibinfo {volume} {150}},\ \bibinfo
  {pages} {052124} (\bibinfo {year} {2009})}\BibitemShut {NoStop}%
\end{thebibliography}
%

\end{document}